\documentclass[10pt,superscriptaddress,nofootinbib,notitlepage,twocolumn,pra]{revtex4-2}

\usepackage{amsthm,amsmath,amssymb}
\usepackage{mathrsfs}
\usepackage{appendix}
\usepackage{amstext}
\usepackage{graphicx}
\usepackage{url}
\usepackage{float}
\usepackage{bm}
\usepackage{ifthen}
\usepackage[usenames,dvipsnames]{color}
\usepackage{mathrsfs}
\usepackage[colorlinks=true, citecolor=blue, urlcolor=black]{hyperref}
\usepackage{dcolumn}
\usepackage{natbib}
\usepackage{booktabs}
\usepackage{hyperref}
\usepackage{color}
\usepackage{multirow}
\usepackage{mathtools}
\usepackage{lipsum}
\usepackage{amssymb}

\newcommand{\bra}[1]{\mbox{$\left \langle #1 \right|$}}
\newcommand{\ket}[1]{\mbox{$\left| #1 \right\rangle$}}
\newcommand{\etal}{\mbox{$et$ $al$. }}

\begin{document}
	
	\title{Breaking universal limitations on quantum conference key agreement without quantum memory}
	
    \author{Chen-Long Li}\thanks{These authors contributed equally to this work}
	\affiliation{National Laboratory of Solid State Microstructures and School of Physics, Collaborative Innovation Center of Advanced Microstrucstures, Nanjing University, Nanjing 210093, China}
	\author{Yao Fu}\thanks{These authors contributed equally to this work}	
	\affiliation{Beijing National Laboratory for Condensed Matter Physics and Institute of Physics, Chinese Academy of Sciences, Beijing 100190, China}
	\author{Wen-Bo Liu}
	\author{Yuan-Mei Xie}
	\author{Bing-Hong Li}
	\author{Min-Gang Zhou}
	\affiliation{National Laboratory of Solid State Microstructures and School of Physics, Collaborative Innovation Center of Advanced Microstrucstures, Nanjing University, Nanjing 210093, China}
	\author{Hua-Lei Yin}\email{hlyin@nju.edu.cn}
	\author{Zeng-Bing Chen}\email{zbchen@nju.edu.cn}
	\affiliation{National Laboratory of Solid State Microstructures and School of Physics, Collaborative Innovation Center of Advanced Microstrucstures, Nanjing University, Nanjing 210093, China}

	\begin{abstract}
	    Quantum conference key agreement is an important cryptographic primitive for future quantum network.
		Realizing this primitive requires high-brightness and robust multiphoton entanglement sources, which is challenging in experiment and unpractical in application because of limited transmission distance caused by channel loss. 
		Here we report a measurement-device-independent quantum conference key agreement protocol with enhanced transmission efficiency over lossy channel.
		With spatial multiplexing nature and adaptive operation, our protocol can break key rate bounds on quantum communication over quantum network without quantum memory.
		Compared with previous work, our protocol shows superiority in key rate and transmission distance within the state-of-the-art technology.
		Furthermore, we analyse the security of our protocol in the composable framework and evaluate its performance in the finite-size regime to show practicality.
		Based on our results, we anticipate that our protocol will play an indispensable role in constructing multipartite quantum network.
	\end{abstract}
	
	\maketitle
	\section{Introduction}
	
	Using quantum physics to process information and building a network with quantum nature in a connected world has established various benefits. 
	Quantum computers offer algorithm speedups~\cite{arute2019quantum,zhong2020quantum,zhong2021phase,wu2021strong}, which are advantageous in interdisciplinary fields such as machine learning~\cite{liu2021rigorous,zhou2022experimental}.
	Quantum communication, especially quantum key distribution and entanglement-assisted point-to-point communication, provides information-theoretic security~\cite{gisin2002quantum, bennett1992communications}.
	In addition, network protocols including blind quantum computation~\cite{broadbent2009universal,stefanie2012demonstration}, distributed quantum computation~\cite{buhrman2003distributed}, quantum secret sharing~\cite{hillery1999qss,Gu2021differential,jia2021differential}, and quantum conference key agreement (QCKA)~\cite{chen2007multi,Cao2021coherent,zhao2020phase,Li2021finite,cao2021high,alasdair2022continuous} have emerged as indispensable building blocks for multiuser applications as well.
	
	Conference key agreement is a cryptographic primitive that shares information-theoretic secure keys among more than two authenticated users for group encryption and decryption~\cite{chen2007multi}.
	This classical cryptographic primitive is vulnerable and no longer secure in the face of eavesdroppers with quantum resources.
	Multiple quantum key distribution links~\cite{bennett1984proceedings,ekert1991quantum,hwang2003decoy,lo2005decoy,wang2005decoy,lo2012measurement,braunstein2012side,lucamarini2018overcoming,liu2021homodyne,xie2022breaking,zeng2022mode} can be directly applied to protect against quantum eavesdroppers.
	However, repetitive use of quantum key distribution links restricts the communication efficiency in a fully connected quantum network.
	Alternatively, multipartite entangled states can be used to realize QCKA for achieving a genuine advantage over the point-to-point quantum communication protocols~\cite{epping2017multi}. 
	Several experimental works on multipartite quantum communication and distribution of the Greenberger-Horne-Zeilinger (GHZ) entanglement~\cite{greenberger1989bell,mermin1990extreme} have been demonstrated~\cite{tittel2005experimental,schmid2005experimental,chen2005experimental,gaertner2007experimental,erven2014experimental,massimiliano2021experimental}.
	Nevertheless, these works remain quite unpractical due to their low key rates and fragility of entanglement resources.
	To avoid requiring entanglement preparation beforehand, a scheme for distributing the postselected GHZ entanglement~\cite{fu2015long} was proposed which combined the decoy-state~\cite{hwang2003decoy,lo2005decoy,wang2005decoy} and measurement-device-independent (MDI) idea~\cite{lo2012measurement,braunstein2012side}.
	However, with the increase in the number of users, this protocol is limited in terms of long-distance deployment due to universal limitations on channel loss~\cite{das2021universal}.
	In recent years, various multiparty quantum communication protocols have been proposed and analysed~\cite{Gu2021differential,Li2021finite,cao2021high,Grasselli2019conference} to enhance the key rate performance for long distance deployment.
    Whereas these protocols are measurement device dependent and cannot be directly extended to more than three participants.
	Furthermore, most works analysed the security of QCKA with infinite resources and calculated the secret key rate in the asymptotic limit, while few works consider finite-key effects~\cite{Grasselli2018finite,Grasselli2019conference}.
	
	In all-photonic quantum repeater~\cite{azuma2015all}, cluster states are utilized to demonstrate polynomial scaling of transmission efficiency with distance which is in fact the idea of spatial multiplexing and adaptive operation.
	Similarly, this idea is applied in adaptive MDI quantum key distribution protocol~\cite{azuma2015allqkd}, where both users send multiple single photon states simultaneously to the central relay who subsequently confirms the arrival of photons by applying quantum non-demolition (QND) measurement and pairs the arrived photons adaptively.
    Inspired by the all-photonic quantum repeater~\cite{azuma2015all} and adaptive MDI quantum key distribution~\cite{azuma2015allqkd}, in this work, we propose an MDI-QCKA protocol with the principle of spatial multiplexing and adaptive operation.
	We investigate the performance of our protocol and the result shows it breaks universal limitations on key rate under at least ten users over the network without quantum memory.
	Compared with other existing QCKA protocols, our protocol outperforms under three users in terms of higher key rates and transmission distance which is more than 300 km within experimentally feasible parameter regime.
	Our protocol can be extended to any number of users flexibly and thus fits well in network deployment.
    On the other hand, our protocol is immune to all detection-side attacks because of its MDI nature.
	Furthermore, we establish the security analysis of our protocol in the composable framework and evaluate the performance in the finite-key regime.
	Based on our results, we believe our protocol manifests potential to be an indispensable building block for practical multiparty applications for quantum networks in the future.

	\section{Results}
	
	\subsection{Quantum conference key agreement protocol}
	
	\begin{figure}[tbp!]
		\includegraphics[width=8.5cm]{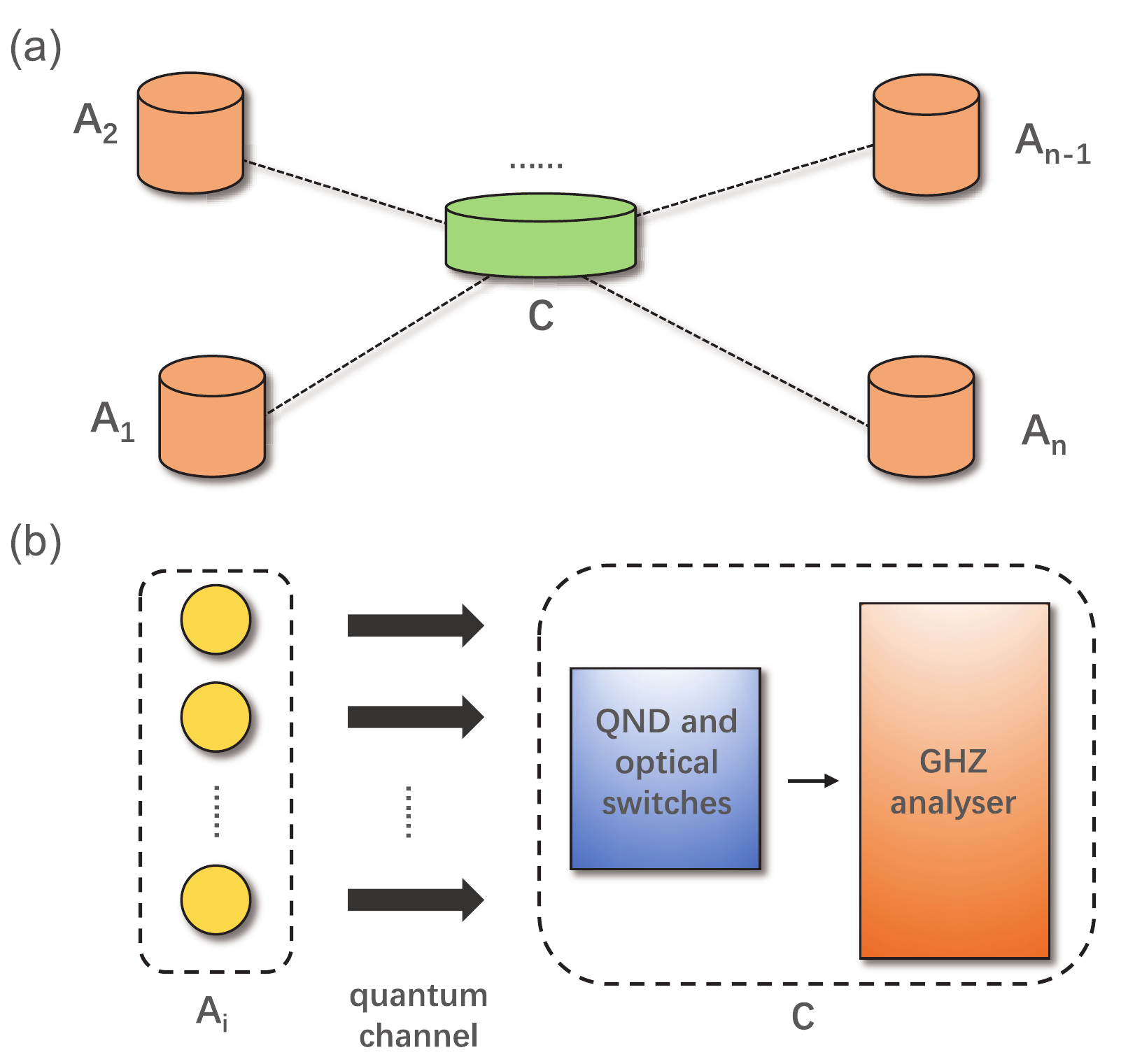}
		\caption{Schematic of our protocol. (a) Network structure of our protocol with $n$ user nodes. (b) Detailed process between $A_i$ and the node $C$. $A_i$ prepares and transmits multiple quantum signals to the central node $C$ through quantum channel. At node $C$ the signals are confirmed by QND measurement and then routed to the GHZ analyser through optical switches. In this process, spatial multiplexing and adaptive operations are used to improve the efficiency of our protocol.
		}\label{qccprotocol}
	\end{figure}
	
	We propose an $n$-party MDI-QCKA protocol to establish postselected GHZ entanglement with spatial multiplexing and adaptive operation.
    Here we denote the $i$th user as $A_i$ $(i=1,...,n)$ and the untrusted central relay as node $C$.
    In Fig.~\ref{qccprotocol} (a), we show the overall structure of our protocol with $n$ user nodes over network.
    In Fig.~\ref{qccprotocol} (b), we show detailed process between $A_i$ and the node $C$.
    Our protocol is described as follows.

	\begin{itemize}
	    \item [$(i)$]
	    All $n$ users $(A_i)_{i=1}^n$ generate $M$ single-photon states that are randomly selected from eigenstates of the $Z$ and $X$ basis.
	    For instance, one selects from $\{\ket{H},\ket{V},(\ket{H}+\ket{V})/\sqrt{2},(\ket{H}-\ket{V})/\sqrt{2}\}$ when using polarization encoding. 
	    $(A_i)_{i=1}^n$ then transmit the $M$ single-photon states to node $C$ simultaneously using spatial multiplexing.
	
	    \item [$(ii)$]
	    Node $C$ performs QND measurements to confirm the arrival of single-photon states from $(A_i)_{i=1}^n$.
	
    	\item [$(iii)$]
    	After the QND measurements, the confirmed photons from every user form a group via optical switches.
	    Node $C$ then performs GHZ projection measurement on the group. 
	    Each user should successfully transmit at least one single photon through QND measurements. 
	    Otherwise this trial is considered to be failed.
	
	    \item [$(iv)$]
	    Node $C$ announces the group information and the GHZ projection results.
    	Each $A_i$ keeps information of states that are successfully projected onto the GHZ state and discards the rest.
	
	    \item [$(v)$]
	    All $n$ communication users $(A_i)_{i=1}^n$ announce their preparing bases for the trials successfully projected onto the GHZ state.
	    If the bases of all parties are the same, the round is kept.
	    The process is repeated until enough rounds have been kept for key generation and parameter estimation.
	
	    \item [$(vi)$]
        The above process is repeated until $m$ trials have been kept for key generation and $k$ trials have been kept for parameter estimation.
        $m$ trials of data for key generation are in the $Z$ basis.
        $k$ trials of data for parameter estimation are in the $X$ basis.

	    \item [$(vii)$]If the test passes, all $n$ users  verify the correctness and proceed with error correction and privacy amplification.
	    If we designate $A_1$ as the conference key reference during error correction, then $A_1$ performs a pairwise information reconciliation with each one of the remaining users.
	    In this process, each one of the remaining users computes a guess of $A_1$'s raw key.
	    If the check passes, they obtains the final secret keys.
	\end{itemize}
	 
	\subsection{Security analysis}
	The security of our MDI-QCKA can be generalized directly from the analysis of quantum key distribution~\cite{Grasselli2018finite}.
	Without loss of generality, we designate $A_1$ as the key reference to conduct classical postprocessing.
	In general, $A_1$'s final key \textbf{S}$_1$ can be quantum mechanically correlated with a quantum state held by the adversary.
	We can define the classical-quantum state describing the correlated system of $A_1$'s final key \textbf{S}$_1$ and eavesdropper $E$
	\begin{equation}
		\rho_{\textbf{S}_1,E}=\sum_{\textbf{S}_1}p(\textbf{S}_1)\ket{\textbf{S}_1}\bra{\textbf{S}_1}\otimes\rho^{\textbf{S}_1}_E,
	\end{equation}
	where the sum is over all possible strings and $\rho^{\textbf{S}_1}_E$ is the joint state of eavesdropper given \textbf{S}$_1$.
	
	Ideally a QCKA protocol is secure if it is correct and secret.
	The correctness means every user holds identical bit strings.
	The secrecy requires $\rho_{\textbf{S}_1,E}=\sum_{\textbf{S}_1}\frac{1}{|\textbf{S}_1|}\ket{\textbf{S}_1}\bra{\textbf{S}_1}\otimes\sigma_{E}$, which means the joint system of the eavesdropper is decoupled from $A_1$.
	However, these two conditions can never be met perfectly.
	Therefore, in practice we define an $\epsilon_{c}$-correct and $\epsilon_{s}$-secret QCKA protocol.
	A QCKA protocol is $\epsilon_{c}$-correct if
	\begin{equation}
		\text{Pr}(\exists i\in\{2,...,n\}, \text{ s.t. }\textbf{S}_1\neq\textbf{S}_i)\le\epsilon_c,
	\end{equation}
	where \textbf{S}$_i$ is the final key string of the $i$th user.
	A QCKA protocol is $\epsilon_{s}$-secret if
	\begin{equation}
		p_{\text{pass}}D\left( \rho_{\textbf{S}_1,E},\sum_{\textbf{S}_1}\frac{1}{|\textbf{S}_1|}\ket{\textbf{S}_1}\bra{\textbf{S}_1}\otimes\sigma_{E}\right) \le\epsilon_s.
	\end{equation}
	$D(\cdot,\cdot)$ is the trace distance and $p_{pass}$ is the probability that the protocol does not abort.
	A QCKA protocol is called $\epsilon_{sec}$-secure with $\epsilon_{sec}\le\epsilon_{c}+\epsilon_{s}$ if it is $\epsilon_{c}$-correct and $\epsilon_{s}$-secret.
	
	Following the result of quantum key distribution~\cite{tomamichel2012tight}, the extractable amount of key $l$ for a $\epsilon_{c}$-correct and $\epsilon_{s}$-secret QCKA is
	\begin{equation}
		l=H^{\epsilon}_{\text{min}}(\textbf{Z}|E)-\text{leak}_\text{EC}-\log_2\frac{1}{\epsilon_c\bar{\epsilon}^2}+2,
	\end{equation}
    where $H^{\epsilon}_{\text{min}}(\textbf{Z}|E)$ is the conditional smooth min-entropy characterizing the average probability that the eavesdropper guesses $A_1$'s raw key \textbf{Z}$_1$ correctly using optimal strategy and leak$_\text{EC}$ is the amount of information leakage of error correction.
    $\epsilon$ and $\bar{\epsilon}$ are positive constants proportional to $\epsilon_{s}$.
	In a realistic scenario, following previous work~\cite{Grasselli2018finite}, the computable key length of our QCKA protocol is
	\begin{equation}\label{QCKAlength}
		l=m\left[q-h(E_X+\mu(E_X,\epsilon'))\right]-\text{leak}_{\text{EC}}-2\log_2\frac{1}{2\bar{\epsilon}},
	\end{equation}
	where
        $\mu(\lambda,\epsilon)=\frac{\frac{(1-2\lambda)AG}{m+k}+\sqrt{\frac{A^2G^2}{(m+k)^2}+4\lambda(1-\lambda)G}}{2+2\frac{A^2G}{(m+k)^2}}$
        with $\lambda$ being the error rate observed in parameter estimation, $A=\max\{m,k\}$ and $G=\frac{m+k}{mk}\ln\frac{m+k}{2\pi mk\lambda(1-\lambda)\epsilon^2}$.
    $q$ is the preparation quality quantifying the incompatibilities of two measurements.
	Detailed proof of the computable key length is shown in Supplementary Note 1.
		
	\subsection{Numerical simulation}
	
	\begin{figure}[tbp!]
		\includegraphics[width=8.5cm]{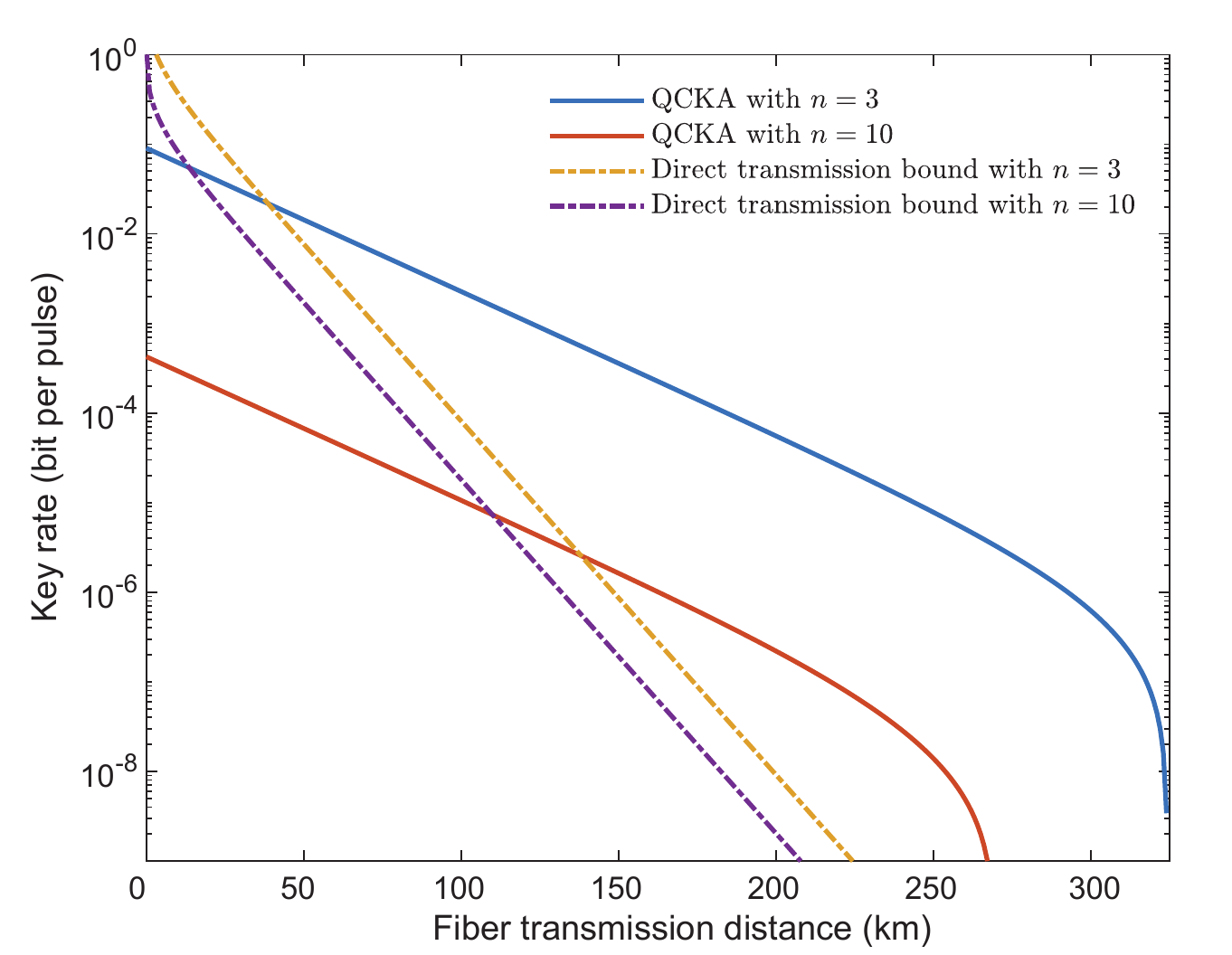}
		\caption{Key rates of QCKA from our protocol and direct transmission bounds. We show key rates of our protocol and corresponding bounds under different numbers of communication parties ($n=3, 10$ from top to bottom). In the figure, key rates of our protocol and bounds are plotted with solid and dash-dotted lines, respectively. The fiber transmission distance denotes the distance between any $i$th party and the central relay.
		}\label{qcc_vs_bound}
	\end{figure}
	
	Before analysing the performance of our protocol, we discuss the universal limitations on quantum communication over network and provide a benchmark for our protocol.
	
	For point-to-point protocols, a fundamental upper limit on the secret key rate over a lossy optical channel not assisted by any quantum repeater is given by $\log_2(\frac{1+\eta}{1-\eta})$ with $\eta$ being the transmissivity between two users~\cite{takeoka2014fundamental}.
	A general methodology allowing to upperbound the two-way capacities of an arbitrary quantum channel with a computable single-letter quantity was devised in~\cite{pirandola2017fundamental}, where the maximum rate achievable by any optical implementation of point-to-point quantum key distribution is given by $-\log_2(1-\eta)$ for the lossy channel.
	For quantum communications over network scenarios, bounds have also been established under different scenarios~\cite{pirandola2019end,pirandola2020general}.
	Furthermore, Das \etal provided a unifying framework to upperbound the key rates of both bipartite and conference settings with different scenarios~\cite{das2021universal}.
    
    In our work, to investigate the performance of our protocol, we consider a rate benchmark in a case where the untrusted central node is removed and all $n$ users are linked by a star network similar to that in Ref.~\cite{Grasselli2019conference}. 
    In such scenario, one user is selected and he performs quantum key distribution with every other $n-1$ users to establish bipartite secret keys with the same length due to the network symmetry.
    The selected user can encode the conference key with the established keys to conduct conference key agreement protocol.
    According to the secret-key capacity, the asymptotic rate is $-\log_2(1-\eta)$ with $\sqrt{\eta}$ being the transmissitivity of the channel from any $i$th user to the central relay.
    Therefore, in this scenario, the key rate is bounded by $\frac{-\log_2(1-\eta)}{n-1}$.
    It should be noted that the above scenario does not necessarily yield the highest key rate in QCKA.
    We denote this bound as the direct transmission bound and use it as a benchmark to evaluate our protocol.
	
	\begin{figure}[tp!]
		\includegraphics[width=8.5cm]{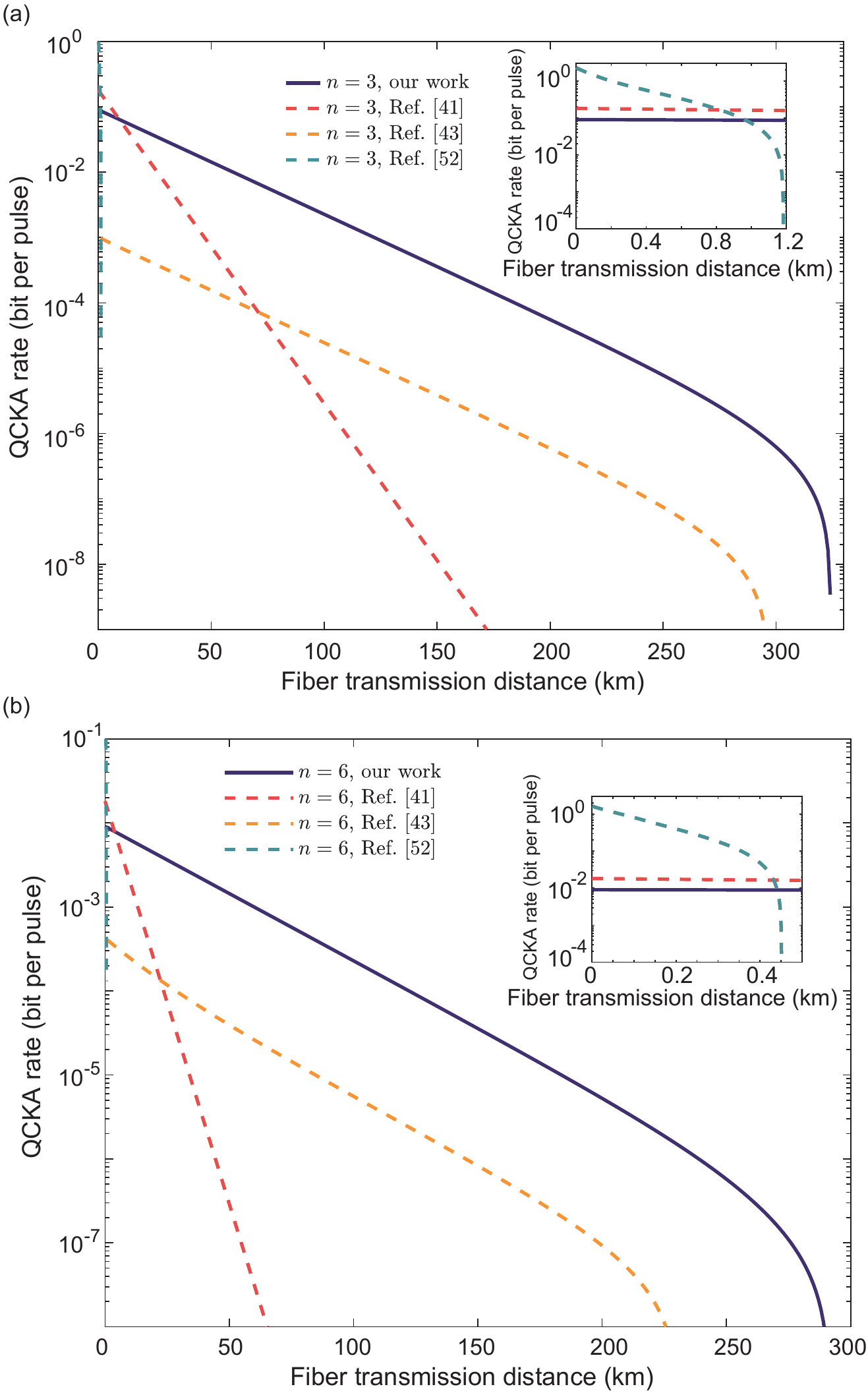}
		\caption{Comparison between key rates of QCKA from our work, MDI-quantum cryptographic conferencing~\cite{fu2015long},  conference key agreement with single-photon interference~\cite{Grasselli2019conference}, and MDI star-network module~\cite{ottaviani2019modular}. We plot the key rates of the protocols when (a) $n=3$ and (b) $n=6$. Different colors are used to denote different communication protocols. For a clear view, the key rate of MDI star-network module~\cite{ottaviani2019modular} is depicted in the inset by restricting the transmission distance to (a) 1.2 km and (b) 0.5 km. The fiber transmission distance denotes the distance between any $i$th party and the central relay.
		}\label{figqcc}
	\end{figure}
	
	In the asymptotic limit, the key rate of QCKA is given by~\cite{fu2015long}
	\begin{equation}\label{qckarate}
		R_{\text{QCKA}}=Q_Z\left[ 1-\text{max}\{h(E_Z^{1,2}),...,h(E_Z^{1,n})\}-h(E_X)\right] ,
	\end{equation}
	where $Q_Z$, is the gain of the $Z$ basis since QCKA generates keys using data from the $Z$ basis.
	$\left\lbrace E_Z^{1,i},i=2,...,n\right\rbrace $ are marginal error rates, which describe the bit error rates between the first user and the $i$th user. 
	$E_X$ is the phase error rate.
	Without loss of generality, here we designate the first user as key reference. 
	
	The gain $Q_Z$ is defined as the efficiency of successful GHZ projection.
	Specifically, we have $Q_{Z}=\frac{\bar{N}}{M}$, where $\bar{N}$ is the average number of postselected GHZ entangled states formed by successfully transmitted photons using $M$ multiplexing.
	Intuitively, if we consider $M$ multiplexing and the total efficiency from any $i$th user to the central node $\eta_t$ including loss and success probability of GHZ projection, then $\bar{N}\sim M\eta_t$.
	Therefore, we have $Q_Z\sim\eta_t$.
	The approximate relation can be converted to an equation $Q_Z=\eta_t$ under the asymptotic limit ($M\rightarrow\infty$). 
	We prove this equation when $n=3$ in Supplementary Note 2.
	To guarantee that more than one postselected GHZ entangled state is generated on average, the number of multiplexing should satisfy $M\ge \eta_t^{-1}$, which implies that $\bar{N}\sim M\eta_t\ge1$.

	In Fig.~\ref{qcc_vs_bound}, we plot the key rates of our QCKA as well as direct transmission bounds as a function of distance between any $i$th party and the central relay with different numbers of communication parties.
    The experimental parameters used in the numerical simulation is presented in Methods.
    Here we consider a symmetric structure where the distance between any user to the central relay is equal.
	We present key rates and bounds with $n=3,10$ users from top to bottom using solid and dash-dotted lines respectively.
	From the simulation results, our protocol overcomes the direct transmission bounds, which stems from the spatial multiplexing and adaptive operations of our protocol. 
	Regardless of the increasing number of communication parties, a polynomial scaling of efficiency with distance can be realized while the bounds attenuate greatly as $n$ increases.
	One can also notice that the key rates of our protocol decrease with increasing $n$ due to the higher error rate when there are more users.
	
	To further investigate the performance of our work, we evaluate the key rate of our protocol and that of other preceding quantum communication protocols over quantum network under the same experimental parameters.
	In Fig.~\ref{figqcc}, we plot the key rate of our QCKA protocol, MDI-quantum cryptographic conferencing~\cite{fu2015long}, MDI star-network module~\cite{ottaviani2019modular}, and conference key agreement with single-photon interference~\cite{Grasselli2019conference} under $n=3$ and $n=6$.
	Our work can reach more than 300 km when $n=3$ and more than 290 km when $n=6$, which shows the capability of intercity scale deployment.
    Therefore, the advantage of our work remains as $n$ grows larger.
	For MDI-quantum cryptographic conferencing, since the gain attenuates exponentially with increasing $n$, the key rate decreases in a similar way.
	The key rate of conference key agreement with single-photon interference shows a performance approximate to that of our work.
	However, the conference key agreement with single-photon interference requires each party to prepare an entangled state $\ket{\phi} = \sqrt{q}\ket{0}\ket{0}+\sqrt{1-q}\ket{1}\ket{1}$ where $q$ is a parameter to be optimized in simulation.
	Preparing such entangled state is quite challenging within available technology.
	A single MDI star-network module can only reach approximately 1 km as shown in the inset of Fig.~\ref{figqcc}.
	Therefore, such modules should be linked together to achieve constant high-rate secure communication over long distances.
 
    To make a comprehensive comparison between different protocols, as shown in Table~\ref{tab_comparison}, we present a table comparing the aforementioned QCKA protocols in different aspects.
    To be specific, we present the longest transmission distance and corresponding key rates of different protocols under $n=3,6,10$ in the first six rows.
    In the last five rows, we compare the other five different aspects including measurement device independence, requirements on entanglement resource, phase stabilization, QND measurement, and whether the protocol is analyzed in the finite-size regime.
    From the table, one can observe that our QCKA protocol shows an advantage in the longest transmission distances and the corresponding key rates.
    In terms of security, all of the aforementioned protocols are measurement-device-independent.
    Only the conference key agreement with single-photon interference~\cite{Grasselli2019conference} requires entanglement resources to conduct the protocol.
    Our protocol and MDI quantum cryptographic conferencing~\cite{fu2015long} avoid requirements for phase stabilization.
    However, QND measurement is needed to confirm the arrival of transmitting photons in our protocol, which is still challenging in experiment.
    Except for MDI quantum cryptographic conferencing~\cite{fu2015long}, other protocols have been analyzed in the finite-size regime.

	\begin{figure}[tbp!]
		\includegraphics[width=8.5cm]{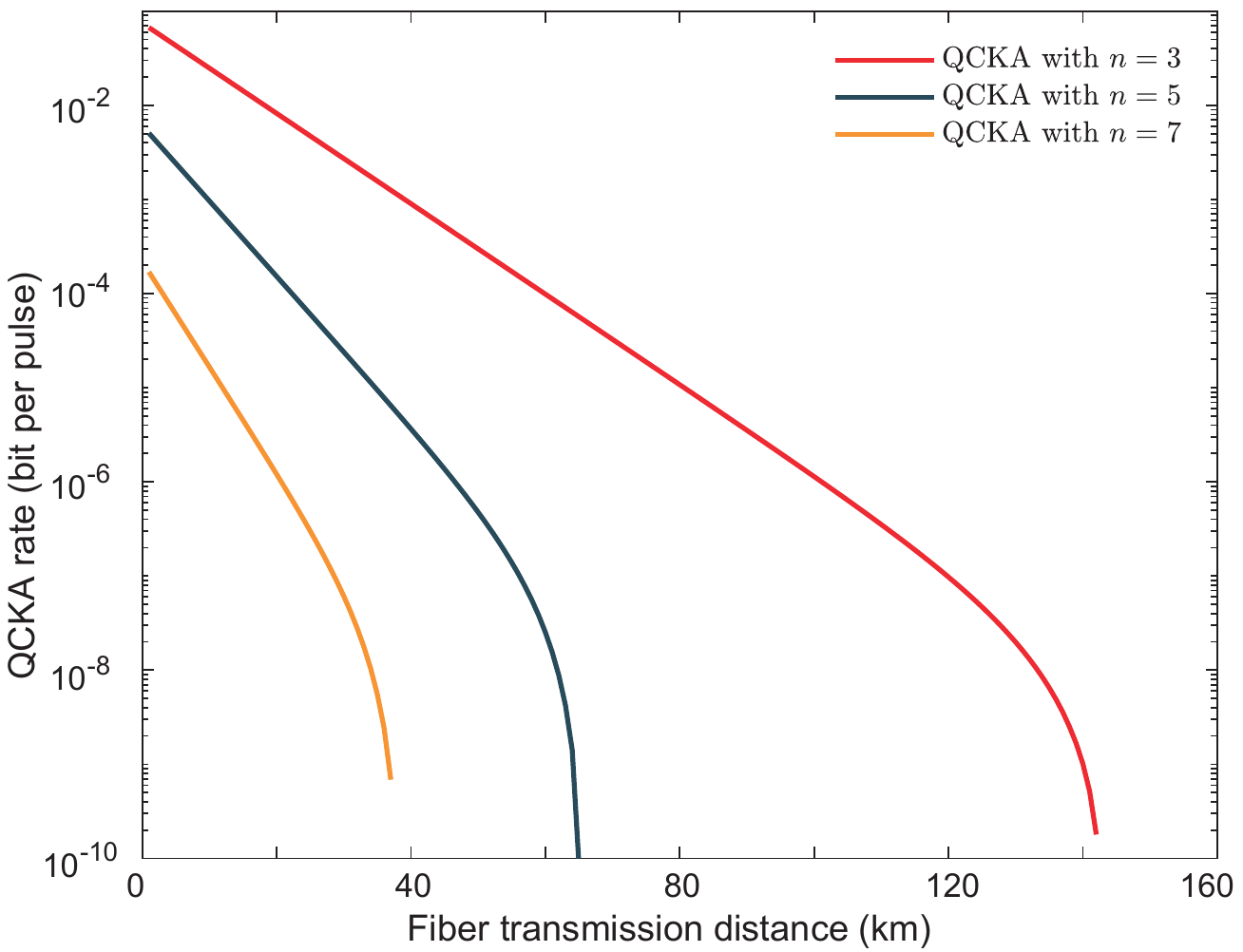}
		\caption{Secret key rate of our QCKA as a function of distance in the finite-size regime. We consider the secret key rate of our QCKA with $n=3,5,7$ shown in different colors. In this simulation, we fix the total number of signals to be $10^{12}$. The fiber transmission distance denotes the distance between any $i$th and the central relay.
		}\label{qcka_dis_finite}
	\end{figure}

	We investigate the performance of our QCKA protocol in the finite-size regime with the same parameters introduced in the asymptotic scenario.
	Furthermore, we fix $\epsilon_{c}=10^{-15}$ corresponding to a realistic hash tag size in practice~\cite{renner2008security}.
    We also fix the total number of signals $L$ to be $10^{12}$.
    Then the number of trials used for key generation can be calculated as $m=p^n\cdot Q_Z\cdot L$, with $p$ the probability of choosing the $Z$ basis which can be optimized to maximize the key rate $l/L$.
	In our protocol, we assume the error correction leakage to be $\text{leak}_{\text{EC}}=fmh(E_Z^{A_1A_i})+\log_2\frac{2(n-1)}{\epsilon_{c}}$~\cite{Grasselli2018finite}, where we set $f=1.1$ and $E_Z^{A_1A_i}$ is the marginal error rate between $A_1$ and $A_i$.
	Then following Eq.~(\ref{QCKAlength}) we can obtain the result in the finite-size regime.
	
	In Fig.~\ref{qcka_dis_finite}, we plot the secret key rate of our QCKA protocol as a function of the distance between any $i$th user and the GHZ analyser.
	In Fig.~\ref{qcka_dis_finite}, we can view that our QCKA protocol can reach more than 140 km, 60 km, and 40 km when $n=3,5,7$, respectively.
	The results are meaningful to practical deployment of an intra- or inter- city quantum network.
	Therefore, we can anticipate our protocol to be essential in the network applications and indispensable for the construction of a connected quantum world.

    \begin{table*}[tp!]
        \setlength{\tabcolsep}{10pt}
		\caption{Comparison between our QCKA and other protocols. In the first six rows we present the longest transmission distance and corresponding key rates of different protocols under $n=3,6,10$. In the last five rows, we compare the other five different aspects including measurement device independence, requirements for entanglement resource, phase stabilization, QND measurement, and whether the protocol is analyzed in finite-size regime.}
		\begin{tabular}{ccccc}
			\hline
			\hline
			&Our QCKA&Ref.~\cite{fu2015long}&Ref.~\cite{Grasselli2019conference}&Ref.~\cite{ottaviani2019modular}\\
			\hline
			Longest transmission distance ($n=3$)&324 km&324 km&296 km&1.18 km\\
            Corresponding key rate (bit/pulse) ($n=3$)&$3.4125\times10^{-9}$&$2.9255\times10^{-19}$&$4.568\times10^{-10}$&$2.9418\times10^{-5}$\\
            Longest transmission distance ($n=6$)&292 km&292 km&231 km&0.45 km\\
            Corresponding key rate (bit/pulse) ($n=6$)&$3.0994\times10^{-9}$&$3.254\times10^{-32}$&$9.6676\times10^{-10}$&$1.2962\times10^{-4}$\\
            Longest transmission distance ($n=10$)&270 km&270 km&148 km&0.26 km\\
            Corresponding key rate (bit/pulse) ($n=10$)&$2.3384\times10^{-10}$&$8.5914\times10^{-49}$&$3.1703\times10^{-9}$&$2.1264\times10^{-4}$\\
			Is measurement device independent.&\checkmark&\checkmark&\checkmark&\checkmark\\
            Requires entanglement resource.&$\times$&$\times$&\checkmark&$\times$\\
            Requires phase stabilization.&$\times$&$\times$&\checkmark&\checkmark\\
            Requires QND measurement.&\checkmark&$\times$&$\times$&$\times$\\
            Has finite-key analysis.&\checkmark&$\times$&\checkmark&\checkmark\\
			\hline
			\hline
		\end{tabular}	\label{tab_comparison}
	\end{table*}

	\section{Discussion}
 
	In this work, we report an MDI-QCKA protocol for quantum network application.
	We analyse the security of the QCKA protocol with composably secure framework and provide a computable key length in the finite-size regime.
	The performance of the QCKA protocol under the GHZ analyser based on linear optical elements~\cite{pan1998greenberger} is investigated.
	Compared with the direct transmission bound of quantum communication over quantum network, our protocol shows great potential in deploying in large-scale quantum network~\cite{cacciapuoti2020quantum,jessica2022quantum}.
	We also show superiority of our work by directly comparing the key rate of our work with those of previous works in multiparty quantum communication~\cite{fu2015long,ottaviani2019modular,Grasselli2019conference}.
	Based on the results of this work, we can anticipate a wide usage of our work in multiparty applications of secure quantum network.
	
	Here we remark on possible directions for future work.
	We have investigated our protocol under a model consisting of single photon sources, QND measurements, optical switches, and the GHZ analyser based on linear optical elements. 
	Further study can be conducted on evaluating the performance of our protocol using other techniques.
	For instance, we investigate our protocol with the GHZ analyser based on linear optical elements which can only identify two of $n$ GHZ states.
	Our protocol can be improved by utilizing the complete GHZ analyser which can identify all $2^n$ GHZ states, such as GHZ state analysis taking into account nonlinear processes~\cite{qian2005universal,Xia2014complete} or entangled-state analysis for hyperentangled photon pairs~\cite{sheng2010complete,liu2015generation}.
	On the other hand, in step $(iii)$ of our protocol, large scale optical switches are needed to route the photons into the GHZ analyser, which may affect the transmittance and cause unwanted loss.
	Thus, future effort should be made towards realizing the protocol with reduced scale optical switches and one possible way is utilizing a Hadamard linear optical circuit together with single-mode on/off switches~\cite{azuma2015allqkd}.
	Investigation of the robustness of our protocol with the existence of multiple-photon components and imperfections in experimental setups should be conducted.
	Techniques in MDI quantum key distribution~\cite{zhou2016making,GU2022experimentalMDI} can be applied in our QCKA to further improve practicality. 
	As we have stated, the all-photonic scheme utilizes cluster states to realize a polynomial scaling with distance which is in fact a result of spatial multiplexing. 
	Therefore, with such spatial multiplexing idea, we can develop other protocols apart from quantum communication with enhanced efficiency.
	In addition, our work can be further developed to give anonymity to users~\cite{grasselli2022anonymous} over quantum network for more complex application scenarios.
	
	\section{Methods}
	
	\subsection{Experimental parameters used in numerical simulation}
	
	In numerical simulation, we use efficiency $\eta_{\text{sps}}$ to describe the probability of the single photon source generating single photons and set $\eta_{\text{sps}}=0.9$~\cite{christensen2013detection}.
	We consider the GHZ analyser based on linear optical elements~\cite{pan1998greenberger} capable of identifying two of the $n$-particle GHZ states.
	We present the working details of the analyser in Supplementary Note 3.
	Photons travel through optical fiber channels whose transmittance is determined by $\sqrt{\eta_{\text{channel}}}=\exp\left(-\frac{l}{l_{\text{att}}} \right)$, where the attenuation distance $l_{\text{att}}=27.14$ km and $l$ is the distance from any $i$th user to the GHZ analyser.
	QND measurements are required to confirm the arrival of photons and the success probability of QND measurements is denoted by $p_{\text{QND}}$.
	To simplify the simulation, we consider a QND measurement for a single photon based on quantum teleportation~\cite{kok2002single}, which uses the fact that the teleportation fails when the incoming state is the vacuum state.
    The QND measurement scheme consists of a Bell state measurement module based on linear optical elements and a parametric down-converter, which we expect is feasible in implementations.
    With ideal parameters we have $p_{\text{QND}}=1/2$.
    Furthermore, with the theoretical and experimental advances in QND measurement of single photons~\cite{distante2021detecting,anderson2022quantum,jiao2022quantum}, we expect the implementation of our protocol to be easier and more efficient in the foreseeable future. 
	Active feedforward technique is needed to direct the arrived photons to the GHZ analyser via optical switches.
	We assume the active feedforward costs time $\tau_a=67$ ns~\cite{ma2011experimental}, which is equivalent to a lossy channel with the transmittance $\eta_a=\exp(-\tau_ac/l_{\text{att}})$, where $c=2.0\times10^8$ $\text{ms}^{-1}$ is the speed of light in an optical fiber.
	Single photon detectors in the GHZ analyser are characterized by an efficiency of $\eta_d=0.93$ and a dark count rate of $p_d=1\times10^{-9}$~\cite{minder2019experimental}, by which we can estimate the success probability of GHZ projection in the $X(Z)$ basis $Q_{X(Z)}^{\text{GHZ}}$.
	Based on the aforementioned assumption on experiment parameters, we analytically estimate the gain with
	\begin{equation}
		Q_{Z}=Q^{\text{GHZ}}_{Z}\cdot p_{\text{QND}}\cdot \eta_a\cdot \eta_{\text{sps}}\cdot \sqrt{\eta_{\text{channel}}}.
	\end{equation}
	See Supplementary Note 4 for the concrete process of estimation of the marginal bit error rates and phase error rate. 
 
\section*{Acknowledgements}
We gratefully acknowledge the supports from the National Natural Science Foundation of China (No. 12274223), the Natural Science Foundation of Jiangsu Province (No. BK20211145), the Fundamental Research Funds for the Central Universities (No. 020414380182), the Key Research and Development Program of Nanjing Jiangbei New Area (No. ZDYD20210101), the Program for Innovative Talents and Entrepreneurs in Jiangsu (No. JSSCRC2021484), and the Program of Song Shan Laboratory (Included in the management of Major Science and Technology Program of Henan Province) (No. 221100210800-02).

$Note$ $Added$.--- After we submitted our work for reviewing, we became aware of a relevant work by Carrara \etal~\cite{carrara2022overcoming}. 
The authors proposed a QCKA protocol using weack coherent pulses and linear optics and proved its security with multiparty decoy-state method.
This protocol can also overcome bounds on the key rate at which conference keys can be established in quantum networks without a repeater.


%

\end{document}